\documentclass[%
 reprint,
showkeys,
showpacs,preprintnumbers,
 amsmath,amssymb,
 aps,
]{revtex4-1}

\usepackage{graphicx}
\usepackage{dcolumn}
\usepackage{bm}

\begin{document}

\newcommand{\CL}{\ensuremath{\mathcal{L}}}
\newcommand{\CT}{\ensuremath{\mathcal{T}}}
\newcommand{\CR}{\ensuremath{\mathcal{R}}}
\newcommand{\CA}{\ensuremath{\mathcal{A}}}
\newcommand{\CY}{\ensuremath{\mathcal{Y}}}
\newcommand{\CF}{\ensuremath{\mathcal{F}}}
\newcommand{\CK}{\ensuremath{\mathcal{K}}}

\newcommand{\defIsoscalarHamiltonian}{
  &H =\sum\limits_\tau ^{(n,p)}\{ \sum\limits_{jm}(E_j-\lambda _\tau )a_{jm}^{\dagger }a_{jm}-\frac 14G_\tau ^{(0)}:(P_0^{\dagger }P_0)^\tau :\nonumber \\
  &-\frac 12\sum\limits_{\lambda \mu }\kappa ^{(\lambda )}:(M_{\lambda \mu }^{\dagger }M_{\lambda \mu }):\}
}

\newcommand{\defrho}{\rho_j = \frac{1}{\sqrt{2j+1}}\sum_m\langle| \alpha^\dagger_{jm} \alpha_{jm} |\rangle     }


\newcommand{\deferpabcsnumber}{\frac{1}{2}\sum_j (2j+1) \left\lbrace 1 - \frac{(1 - 2\rho_j)(E_j - \lambda)}{\sqrt{(E_j - \lambda)^2 + \Delta^2 }} \right\rbrace = n}
\newcommand{\deferpabcsgap}{\frac{G}{4} \sum_{j} \frac{2j+1}{\sqrt{(E_j-\lambda)^2 + \Delta^2 }} (1 - 2 \rho_j) = 1}
\newcommand{\deferparpa}{\frac{\kappa_\lambda}{2\lambda+1} \sum_{jj'} (1 - \rho_{jj'}) \frac{(f^\lambda_{jj'} u^+_{jj'})^2 (\varepsilon_j + \varepsilon_{j'})}{(\varepsilon_j + \varepsilon_{j'})^2 - \omega_{\lambda i}^2} = 1}
\newcommand{\deferpanorm}{\sum_{jj'} (1 - \rho_{jj'}) [(\psi^{\lambda i}_{jj'})^2 - (\varphi^{\lambda i}_{jj'})^2] = 2}
\newcommand{\deferparho}{\rho_j = \frac{1}{2} \sum_{\lambda i j'} \frac{2\lambda + 1 }{2j+1} (1 - \rho_{jj'}) (\varphi_{jj'}^{\lambda i})^2}






\newcommand{\OddWaveFunction}{\Psi _\nu (JM) = C_{J\nu}\alpha _{JM}^+ +  \sum_{j\lambda i}  D_{j\lambda i}(J\nu )P_{j\lambda i}^{\dagger}(JM) - E_{J\nu}\tilde{{\alpha}}_{JM}-\sum_{j\lambda i} F_{j\lambda i}(J\nu)\tilde{P}_{j\lambda i}(JM) |\rangle}


\newcommand{\EnergyMatrixSimpliefied}{\left( {\begin{array}{*{20}c}
      \varepsilon_J & {V\left( {Jj'\lambda ' i' } \right)} & 0 & { - W\left( {Jj' \lambda ' i' } \right)}  \\
      {V\left( {Jj \lambda i } \right)} & {K_J(j\lambda i | j' \lambda i')} & { W\left( {Jj \lambda i } \right)} & {0}  \\
      0 & { W\left( {Jj' \lambda ' i' } \right)} & -\varepsilon_J & { - V\left( {J j' \lambda ' i' } \right)}  \\
      { - W\left( {Jj \lambda  i } \right)} & {0}  & { - V\left( {Jj \lambda i } \right)} & {-K_J(j\lambda i | j' \lambda i')}  \\

  \end{array} } \right)\left( \begin{gathered}
    C_{J \nu}  \hfill \\
    D_{j'\lambda ' i'}(J\nu)  \hfill \\
    - E_{J\nu}  \hfill \\
    - F_{j' \lambda ' i'}(J\nu)  \hfill \\
  \end{gathered}  \right) }

\newcommand{\EnergyMatrixRHSSimplified}{= \eta _{J\nu}
  \left( {\begin{array}{*{20}c}
      1 & 0 & 0 & 0  \\
      {0} & {1 - \CL^{*}(Jj\lambda i) } & {0} & {0}  \\
      0 & 0 & 1 & 0  \\
      0 & {0}  & 0 & 1 - \CL^{*}(Jj\lambda i)  \\
  \end{array} } \right)\left( \begin{gathered}
    C_{J \nu}  \hfill \\
    D_{j'\lambda ' i'}(J\nu)  \hfill \\
    - E_{J\nu}  \hfill \\
    - F_{j'\lambda ' i'}(J\nu)  \hfill \\
  \end{gathered}  \right)}


\newcommand{\VFINALSIMPLIFIED}{-\frac{1}{\sqrt{2}}[1-\rho_j + \CL^{*}(Jj\lambda i)]\Gamma(Jj\lambda i)}

\newcommand{\WONEPART} {\frac{\pi_{\lambda}}{\pi_{J}} \varepsilon_{J} \rho_{j} \varphi_{Jj}^{\lambda i}}
\newcommand{\WFINALSIMPLIFIED}{     \WONEPART  -\frac{1}{4} [ 1 - \rho_j + \CL^{*}(Jj\lambda i)]  \frac{\pi_\lambda}{\pi_J}\sum_{i_{1}} \CA(\lambda i_{1} i) \varphi_{Jj}^{\lambda i_{1}} }

\newcommand{\KFINALSIMPLIFIED}{ \delta_{jj^{\prime}}\delta_{\lambda\lambda^{\prime}}\delta_{ii^{\prime}}[1-\rho_{j} + \CL^{\ast}(Jj\lambda i)](\varepsilon_{j} + w_{\lambda i})  \nonumber\\
  & - \delta_{jj^{\prime}}\delta_{\lambda\lambda^{\prime}}\delta_{ii^{\prime}}(1+\CL(Jj\lambda i))\frac{1}{4}\sum_{\i_{1}}\CA(\lambda ii_{1})\CL^{\ast}_{J|j}(j\lambda i|j\lambda i_{1}) }

\newcommand{\defCL}{  \CL^{*}_{J|j'}(j\lambda i | j' \lambda'i' ) = \pi_{\lambda\lambda'} \sum_{j_{1}} ( 1 - \rho_{j_{1}j'}) \psi_{j_1 j}^{\lambda' i'} \psi_{j_1 j'}^{\lambda i} \left\{ {\begin{array}{*{20}c}
    {j' } & {j_{1} } & \lambda   \\
    {j } & J & \lambda'  \\
    \end{array} } \right\}}

\newcommand{\defCLSimplified}{  \CL^{*}(Jj\lambda i) = \pi_{\lambda\lambda} \sum_{j_{1}} ( 1 - \rho_{j_{1}j'}) \psi_{1j}^{\lambda i} \psi_{1j}^{\lambda i} \left\{ {\begin{array}{*{20}c}
    {j } & {j_{1} } & \lambda   \\
    {j } & J & \lambda  \\
    \end{array} } \right\}}



\newcommand{\defPhonon}{Q_{\lambda \mu i}^{\dagger}\,=\,\frac 12\sum_{jj^\prime }[\psi_{jj^\prime}^{\lambda i}\,A^{\dagger}(jj^\prime;\lambda \mu)-(-1)^{\lambda -\mu }\varphi _{jj^\prime}^{\lambda i}\,A(jj^\prime ;\lambda -\mu )]}

\newcommand{\defQuasiparticleone}{\alpha _{jm} = u_j a_{jm} - (-)^{j-m}v_j a_{j-m}^{\dagger}}

\title{An Extended Approximation for the Lowest-lying States in Odd-mass Nuclei}

\author{S. Mishev} \email{mishev@theor.jinr.ru}
\author{V.V.Voronov}%
 \email{voronov@theor.jinr.ru}
\affiliation{%
Joint Institute for Nuclear Research \\
  6 Joliot-Curie str. \\
  Dubna 141980, Russia
}%

\date{\today}

\begin{abstract}
An enhanced model, based on the Extended Boson Approximation, for the
lowest-lying states in odd-mass nuclei is presented. Our approach is
built on the Quasiparticle Phonon Model, extending it to take into
account the ground state correlations due to the action of the Pauli
principle more accurately than in the conventional theory. The derived
interaction strengths between the quasiparticles and the phonons in
this model depend on the quasiparticle occupation numbers explicitly
coupling the odd-mass nucleus equations with those of the even-even
core. Within this model we calculated the transition probabilities in
several Te, Xe and Ba isotopes with A$\approx$130.
\end{abstract}

\maketitle


\section{Introduction}

Due to its simplicity and numerous successful applications the Random
Phase Approximation (RPA) \cite{1961_Thouless, 1961_Brown, Rowe68} is
widely considered as a good first approximation to study small
fluctuations in atomic nuclei. However, this simple model enjoys only
a limited success when one needs to describe properties of states from
the lowest part of the spectrum in nuclei remote from the magic
configurations. The Quasi-Boson Approximation (QBA), underlying the
RPA, stimulates a discussion concerning its applicability to the
problem of correctly taking into account the ground state correlations
(GSC) in even-even nuclei. Numerous improvements of this theory with
respect to adding correlations in the ground states of even-even
nuclei have been attempted, as for example in
\cite{Rowe68,LenskeWambach90,Catara94, Providencia68,
  Dukelsky96}. These enhanced models stem from the disregard of the
QBA and are related to more precise inclusion of the Pauli principle
when calculating matrix elements of various operators. An enhanced
version of this approximation, referred to as an Extended RPA (ERPA),
which was proposed a long time ago \cite{Hara} and later developed in
\cite{Jolos,Karadjov}, proved successful in improving the theoretical
results for most measurable quantities near the nuclear ground states
as, for example, the transition charge densities in the interior
region.

In the present work, we follow the ERPA approach, extending it to
provide a refined version of the Quasiparticle Phonon Model (QPM) for
odd-even nuclei \cite{Soloviev, Khuong81,Vdo85, Mish08}. The
interaction strengths between the quasiparticles and phonons in the
presented model depend on the number of quasiparticles in the ground
state. In this way, the core-particle equations couple with the
generalized equations describing the pairing correlations and the
excited vibrational states of the even-even core, thus forming a large
nonlinear system.  This model is applicable to open-shell spherical
and transitional odd-A nuclei where the Pauli principle effects are
becoming essential as the number of nucleons in the unclosed shell
increase.

Our research descends from the studies presented in \cite{Sluys93,
  Mish08}. There it has been shown that the backward amplitudes in the
wave functions of these nuclei play a very important role for better
agreement with the experimentally measured spectroscopic factors and
the properties of the states from the lower part of the energy
spectrum. The theory in the latter papers is based however on the QBA
which we intend to improve by taking into account the action of the
Pauli principle more precisely due to the Extended Boson Approximation
(EBA) \cite{Hara}.

Another widely adopted approach to study odd-A nuclei presents the
Interacting Boson-Fermion Model (IBFM).The IBFM, introduced in
\cite{1979_Iachello} and further extended in numerous papers
(e.g. \cite{AAX}), differs from our approach by that the excited
states of the even-even core nucleus are created by operators of a
pure boson nature. In the IBFM the core-particle interaction depends
on a number of free parameters which are usually fitted to match the
spectrum in the odd-A nucleus. In this respect the QPM is closer to
the interacting shell model where this interaction is derived from the
dynamics of the constituting nucleons.

\begin{figure*}[htb]
  \includegraphics[width=\textwidth]{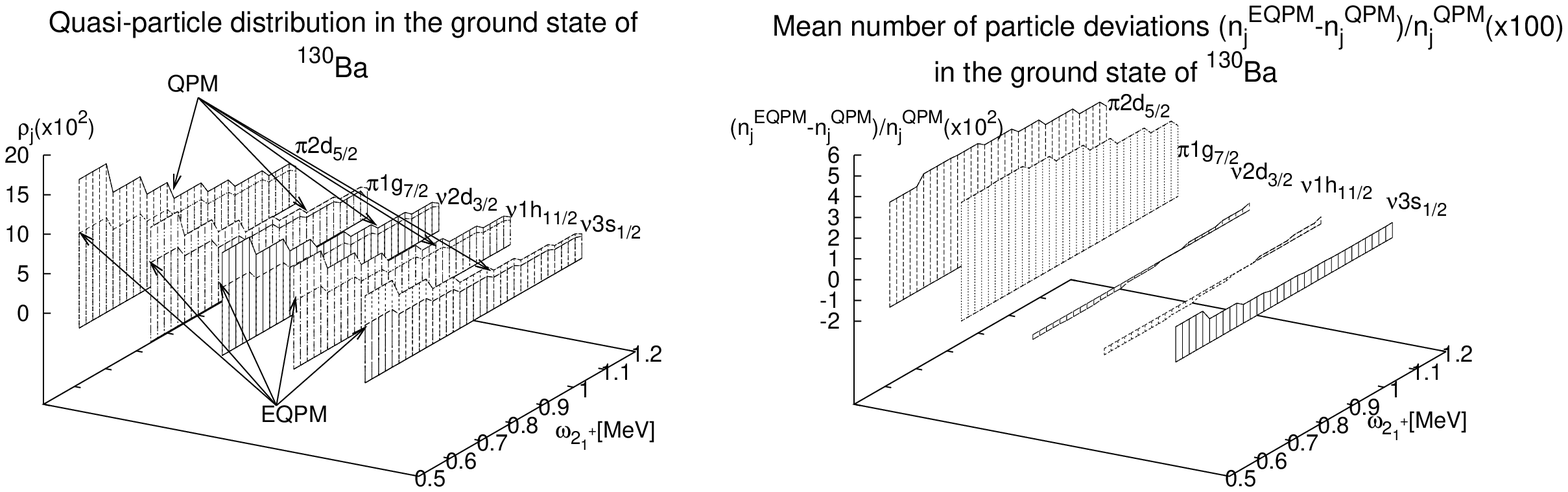}
  \caption{ \label{fig:rho1} Left panel - quasiparticle occupation
    numbers $\rho_{j}\times 100$ in the ground state of $^{130}$Ba
    within a one-phonon QPM and EQPM theory for the sub-shells in the
    valence shell. Right panel - same as in the left panel but for the
    quantities $\left({n_{j}^{EQPM} -
      n_{j}^{QPM}}/{n_{j}^{QPM}}\right)\times 100$, where $n_j$ is the
    number of particles on the level $j$. The quantities in both
    panels are plotted as a function of the first quadrupole phonon's
    energy.}
\end{figure*}

This paper first outlines the QPM and its extension based on the EBA
in even-even nuclei. A comparison between the models built on the QBA
and EBA is established on the basis of the reduced transition
probabilities from the ground to the first $2^{+}$ state in
even-even nuclei with $A \approx 130$. In section \ref{sec:odd-even},
we give the QPM theory for odd-even nuclei emphasizing the effect of
the renormalization on the interaction vertices. Calculations on the
spectroscopic factors and transition probabilities between states in
some odd-even Te,Xe and Ba nuclei, where experimental data are
available, are presented in section \ref{sec:numerics}. Conclusions
are drawn in section \ref{sec:conclusions}.


\section{Even-even nuclei}
\label{sec:even-even}

This section aims to mark the basic building blocks of the QPM and its
EBA extension (EQPM) for one-phonon states. The notations used below
are the same as in \cite{Karadjov} and \cite{Mish08}.

In EQPM one defines the quantities $\rho_{j}$, which are proportional to the quasiparticle occupation numbers in the ground state on the level $j$:

\begin{equation}
  \label{def_rho}
  \defrho,
\end{equation}

where $\alpha$ denotes a quasiparticle ("qp" for short)

\begin{equation}
  \defQuasiparticleone.
\end{equation}

The other key constituent of the theory is the phonon operators ("ph" for short) defined as 
\begin{equation}
  \defPhonon.
\end{equation}

The ground state $|\rangle$ in equation \eqref{def_rho} is the vacuum
state for the phonon operators: $Q_{\lambda \mu i}|\rangle = 0$.

We study the dynamics of nuclear systems governed by the simple
Hamiltonian in the form:

\begin{align}
  \label{eq:Hamiltonian}
  \defIsoscalarHamiltonian .
\end{align}

accounting for the nuclear mean field, the pairing and the isoscalar
multipole-multipole interactions, respectively.

\begin{figure*}
  \includegraphics[width=\textwidth]{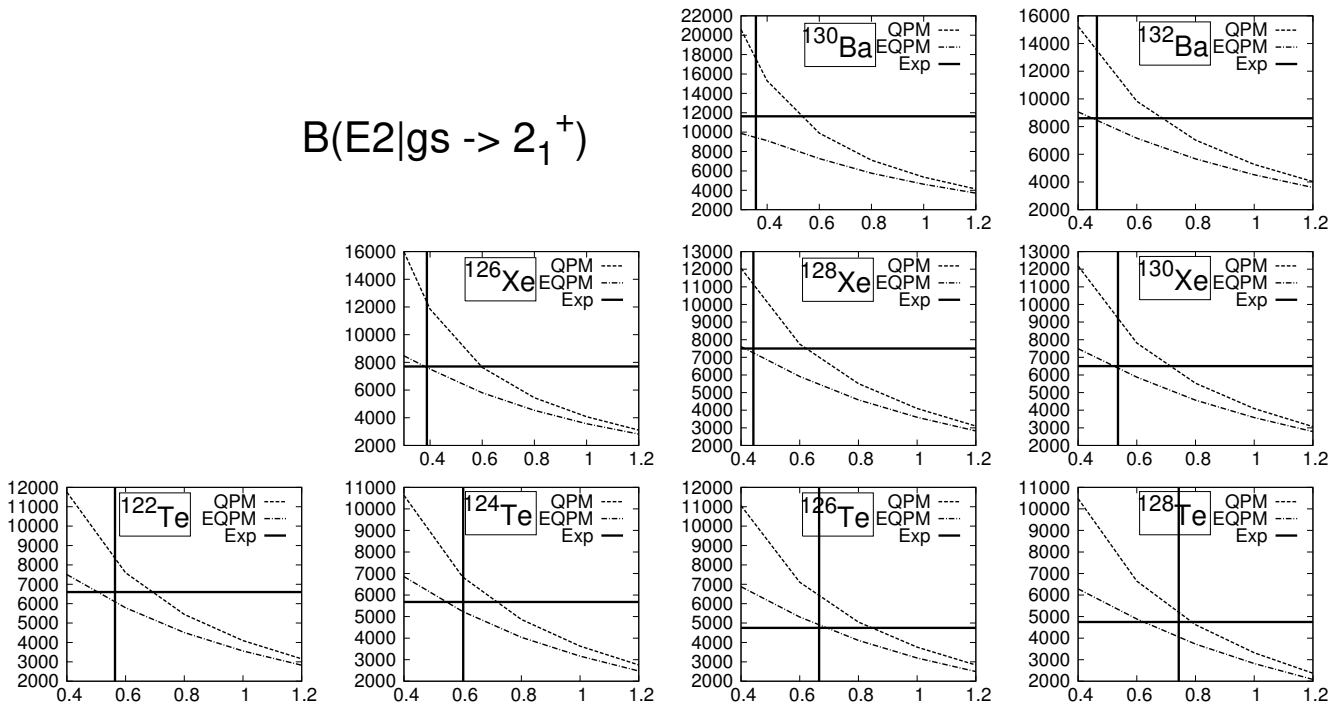}
  \caption{\label{fig:trans_even_even} The reduced transition probabilities $B(E2|gs\rightarrow 2_{1}^{+}$) (in units $e^{2}fm^{4}$) in  several Te, Xe and Ba isotopes plotted against the energy $\omega_{2_{1}^{+}}$ of the first quadrupole phonon. The solid lines represent the experimental energies and transitions. } 
\end{figure*}

If the pairing vibrations are not taken into consideration then one
can obtain \cite{Karadjov} the following modified QPM equations
describing the states in even-even nuclei:

\begin{align}
  & \deferpabcsnumber \label{deferpabcsnumber}\\
  & \deferpabcsgap \\
  & \deferparpa \\
  & \deferpanorm \\
  & \deferparho \label{deferparho}.
\end{align}

The emergence of the factors $(1 - \rho_{jj^\prime})$ takes into
account the blocking effect due to the Pauli principle and requires
one to solve the equations above as a system of coupled equations.

The multipole-multipole interaction strengths $\kappa^{(\lambda)}$ are
treated as free parameters in our study. In the numerical calculations
we kept the quadrupole-quadrupole term only because it gives the
dominant part of the long-range interaction for the determination of
the low-lying states' properties in the nuclei of interest. One way to
fix the parameter $\kappa^{(2)}$ is to have it to reproduce the energy
of the first $2^+$ state ($\omega_{2_1^+}$) . Since a one-to-one
correspondence between $\omega_{2_1^+}$ and $\kappa^{(2)}$ exists, we
show most of the calculated quantities as a function of
$\omega_{2_1^+}$ because its values are more intuitive and closer to
the experiment than the corresponding interaction strength values.

Below we discuss the results obtained within the EQPM for the quasiparticle and particle occupation numbers as well as for the transition probabilities in even-even nuclei.

The differences between the quasiparticle and particle occupation
numbers in $^{130}$Ba as a function of the first quadrupole phonon's
energy within the QPM and EQPM are presented in
Fig. \ref{fig:rho1}. From this figure we see that the smearing of the
Fermi surface increases together with the strength of the field force
(and, correspondingly, $\omega_{2_1}$ decreases). In the right panel
we point out that the relative difference of the particle occupation
numbers calculated within the two model variants can reach up to 5
$\%$, as is the case for the proton sub-shell $2d_{5/2}$.

The transition probabilities in odd-even nuclei are directly linked to the transition probabilities in their corresponding even-even cores as it will be discussed in section \ref{sec:numerics}. We therefore perform a comparative study of the reduced transition probabilities $B(E2|gs\rightarrow 2_1^+)$ in several even-even nuclei within the QPM and EQPM. The transition probabilities in the EQPM are given as

\begin{equation}
  \label{eq:transition_even_even}
  B(E\lambda|g.s. \rightarrow \lambda_i)=\left[\frac{1}{2}\sum_{jj^{\prime}}(1-\rho_{jj^{\prime}})f^{\lambda}_{jj^{\prime}}u^{+}_{jj^{\prime}}g^{\lambda i}_{jj^{\prime}}\right] ^{2}.
\end{equation}

The nuclei presented in Fig. \ref{fig:trans_even_even} were chosen
to be close to spherical ones, having
$E(4_{1}^{+})/E(2_{1}^{+})<2.5$. From this figure we can see that the
blocking effect due to the Pauli principle exerts a large impact on
this measurable quantity. The obvious superiority of the EQPM in this
region serves as a motivation to study odd-even systems with a core
described within the framework of this model. Besides, the transition
charge densities, being related to the reduced transition
probabilities, were studied in \cite{Karadjov}. There it was shown
that the application of the EQPM leads to a better reproduction of the
experimentally measured distributions in the nuclear interior.


\section{Odd-even nuclei}
\label{sec:odd-even}

In our treatment the states in odd-even nuclei are described as mixed states composed of pure quasiparticle and quasiparticle$\times$phonon(qp$\times$ph) states including backward-going amplitudes \cite{Sluys93, Mish08}

\begin{figure*}[htb]
  \includegraphics[width=\textwidth]{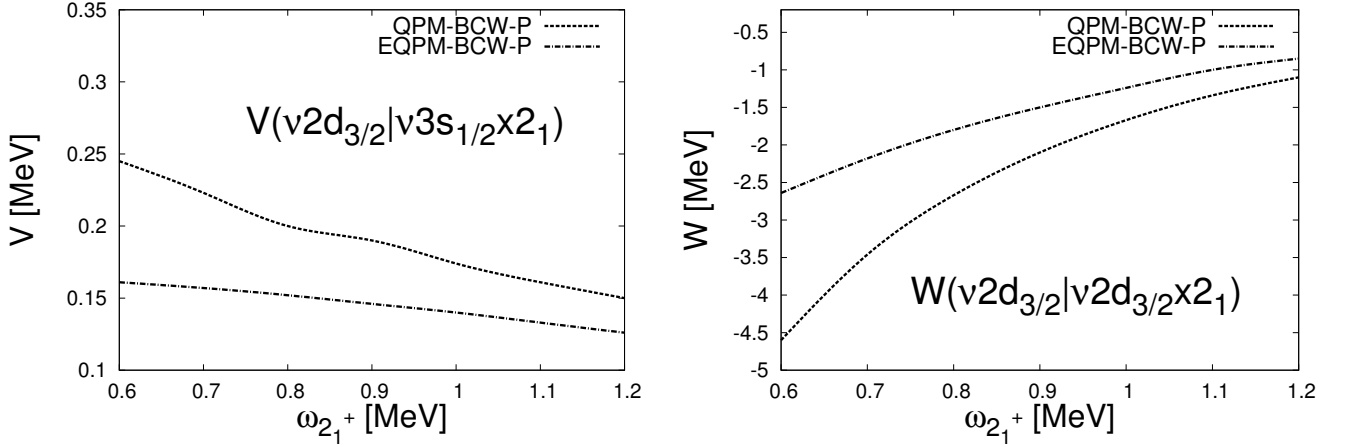}
  \caption{  \label{fig:WVR} The matrix elements $V(\nu 2d_{3/2}|\nu 3s_{1/2} \times 2_{1}^{+} ) $ and $W(\nu 2d_{3/2}|\nu 2d_{3/2} \times 2_{1}^{+})$ in $^{131}$Ba plotted against the energy $\omega_{2_{1}^{+}}$ of the first quadrupole phonon in $^{130}$Ba.}
\end{figure*}

\begin{widetext}
  \begin{equation}
    \label{eq:oddwavefunc}
    \OddWaveFunction,
  \end{equation}
\end{widetext}

where $P^{\dagger}_{j\lambda i}(JM) = [\alpha^{\dagger}_{j}Q^{\dagger}_{\lambda i}]_{JM}$ is the qp$\times$ph creation operator and $\tilde{}$ stands for time conjugation, according to the convention $\tilde{a}_{jm} = (-1)^{j-m}a_{j-m}$.

The structure coefficients from \eqref{eq:oddwavefunc} and the energies of the states in the odd-A nucleus can be obtained by making use of the equation of motion method. Conforming to the relation \eqref{def_rho}, when calculating the matrix elements, we obtain the following generalized eigenvalue problem:

\begin{widetext}
  \begin{align}
    \label{energy_matrix}
    &\EnergyMatrixSimpliefied =  \nonumber \\
    &\EnergyMatrixRHSSimplified.
  \end{align}
\end{widetext}

For conciseness, we provide only the leading terms of the expressions for the matrix elements.

\begin{align}
  & V(Jj\lambda i) = \langle | \lbrace [\alpha_{JM}, H], P^{\dagger}_{j\lambda i}\rbrace|\rangle = \label{eq:V} \\
  & = \VFINALSIMPLIFIED ,  \nonumber
\end{align}
\begin{align}
  & W(Jj\lambda i) = \langle | \lbrace [\alpha^{\dagger}_{JM}, H], \tilde{P}^{\dagger}_{j\lambda i}\rbrace|\rangle = \label{eq:W} \\
  & = \WFINALSIMPLIFIED ,  \nonumber
\end{align}
\begin{align}
  \label{eq:K}
  & K_{J}(j\lambda i|j'\lambda' i') = \frac{1}{2}[I_J(j\lambda i|j'\lambda' i') + I_J(j'\lambda' i'|j\lambda i)]= \\
  & =\KFINALSIMPLIFIED  \nonumber.
\end{align}

For the numerical calculations we used a diagonal approximation for $\CL$ \cite{Khuong81}. Below we list the notations entering into the matrix elements \eqref{eq:V}-\eqref{eq:K}:

\begin{align}
  & I_J(j\lambda i|j'\lambda' i') = \langle| \lbrace P_{j\lambda i}(JM),[H, P^+_{j'\lambda'i'}(JM)] \rbrace |\rangle , \\
  & \defCL , \\
  & \defCLSimplified ,
\end{align}

\begin{align}
  &A(\lambda ii^\prime) = \sum_\tau \frac{X_{\lambda i}(\tau)+X_{\lambda i^\prime}(\tau)}{\sqrt{\CY_{\lambda i}(\tau)\CY_{\lambda i^\prime}(\tau)}}, \\
  &X_{\lambda i}(\tau) = \sum^{\tau}_{jj'}\frac{ (1-\rho_{jj'}) (f^\lambda_{jj'} u_{jj'}^+)^2 \varepsilon_{jj'} }{ \varepsilon^2_{jj'} - \omega_{\lambda  i}^2}, \\
  & \CY_{\lambda i}(p) = \CY_{\lambda i}(n) = \omega_{\lambda i}\sum_{jj'}\frac{ (1-\rho_{jj'}) (f^\lambda_{jj'} u_{jj'}^+)^2 \varepsilon_{jj'} }{ (\varepsilon^2_{jj'} - \omega_{\lambda i}^2)^2}.
\end{align}

 In the limit case $\rho_j = 0$, the problem in \eqref{energy_matrix}
 is brought to the model obtained in \cite{Mish08}. Below we discuss
 the effect of the correlations in the nuclear ground state on the
 behavior of the matrix elements in \eqref{energy_matrix}.

The interaction between the quasiparticles and the phonons will
naturally become stronger when the smearing around the Fermi level
increases. In Fig.\ref{fig:WVR}, the dependence of sample qp-ph
interaction strengths on $\omega_{2_1^+}$ is plotted. The weakening of
this interaction within the extended model, as compared to the
interaction derived within the QBA, is getting more salient as the
ground state correlations increase. It is also worth noting that the
strengths in the backward direction depend not only on the structure
of the phonon state $|\lambda i\rangle$ building the matrix element
$W(Jj\lambda i)$ but also on all other phonons entering into the sum
in the second summand of the rhs of equation \eqref{eq:W} . This
implies that the higher-lying phonon states influence the properties
of the states near the ground state. We estimated that the
contribution of the higher-lying phonons to the quantities
$W(Jj\lambda i)$ can be up to 25$\%$. The diagonal matrix elements
$K_{J}(j\lambda i|j'\lambda' i')$ exhibit a similar effect due to the
second summand of the rhs of \eqref{eq:W}. This sum generates an
energy shift, which can contribute to the appearance of intruder
states in the lower part of the energy spectrum, as discussed in
detail in \cite{Khuong81}.

In our previous paper \cite{Mish08}, it was found out that the
decrease in the energy of the first $2^+$ state leads to a
considerable growth of the quantities $W(Jj\lambda i)$ thus pushing
the first solution very close to the first qp$\times$ph pole. This did
not allow us to correctly reproduce both the properties of the
odd-even nucleus and its even-even core using the same values for the
multipole constants $\kappa_\lambda$ and correspondingly
$\omega_{\lambda_1^{\pi}}$. We noticed that the values of
$\omega_{2_1^+}$ in the even-even core, which let us reproduce the
energies of the lowest part of the spectrum in the odd-even nucleus
with reasonable accuracy, were much higher than their experimental
counterparts. In this regard the weakened interaction between the pure
qp and qp$\times$ph configurations \eqref{eq:V}-\eqref{eq:W} caused by
the quasiparticle blocking should yield better agreement between the
theory and the experiment.

\section{Numerical results}
\label{sec:numerics}

In this section, we present numerical results showing the influence of the backward propagating terms on the spectroscopic factors and the transition probabilities between states in odd-even nuclei using the two approximations - QBA and EBA - giving rise to different variants of the model. The latter are denoted in a similar way as in \cite{Mish08}:
\begin{itemize}
\item {QPM$\_$P - one-phonon model, including Pauli principle corrections (as in \cite{Khuong81}) }
\item {QPM$\_$BCK$\_$P - one-phonon model, including backward amplitudes and  Pauli principle corrections (as in \cite{Mish08}) }
\item {EQPM$\_$BCK$\_$P - one-phonon model, including both backward amplitudes and Pauli principle corrections having a core described within the EQPM.}
\end{itemize}

Below we give technical details on the calculations performed.

\begin{figure}
  \centering
  \includegraphics[width=0.5\textwidth]{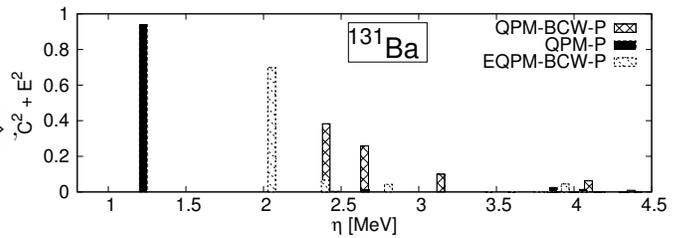}
  \caption{\label{fig:fragmentation} Quasiparticle strength distribution ($C^2 + E^2$) of the state $\nu 2d5/2$ in $^{131}$Ba. The quadrupole-quadrupole interaction strength $\kappa^{(2)}$ is kept constant in the calculations within the three model versions.}
\end{figure}

\begin{figure}
  \includegraphics[width=0.4\textwidth]{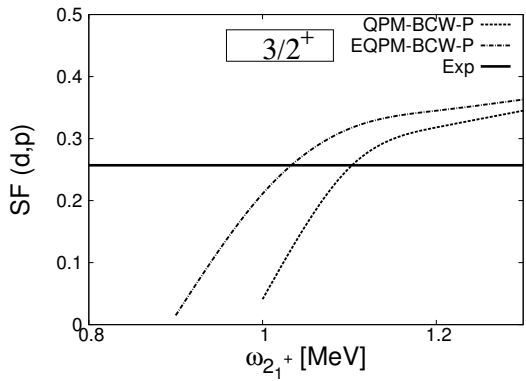}
  \caption{\label{fig:sf1}Spectroscopic factor for the (d,p) reaction in $^{131}$Ba as a function of $\omega_{2_{1}^{+}}$.}
\end{figure}

For simplicity, we employed a Wood-Saxon mean field with parameters
fitted to reproduce the nuclear binding energies. In a similar way,
the pairing strengths $G_\tau$ were obtained to match the odd-even
mass differences in neighboring nuclei(for details see
\cite{Mish08}). We included quadrupole phonons only since the
quadrupole-quadrupole interaction, along with the pairing interaction,
plays a dominant role for the low-lying collective states in even-even
nuclei, as already pointed out in sec.\ref{sec:even-even}. We let the
quadrupole strength $\kappa^{(2)}$, correspondingly
$\omega_{2_{1}^{+}}$, vary and analyze the dependence of the
quantities of interest on $\omega_{2_{1}^{+}}$. The phonons' energy
cutoff is set to 15 MeV. One appealing feature of the QPM and in
particular of the variant described in this paper is that the
interaction strengths between the quasiparticles and phonons depend
only on the parameters describing their internal structure, thereby
introducing no extra degrees of freedom.

From the computational perspective solving the algebraic system \eqref{deferpabcsnumber}-\eqref{deferparho} is a more challenging task than solving the equations of the standard QPM. As an initial approximation to the solution of the coupled problem we take the solutions obtained from the uncoupled equations (i.e. $\rho_{j} = 0$).

We tested the so-developed approximation on several odd-A Te, Xe and
Ba isotopes entering into the transitional region. As it has already
been pointed out in sec.\ref{sec:even-even}, where we investigated the
properties of the corresponding even-even cores, the use of the EQPM
improves the agreement between the results of the calculations and the
experimental data significantly.

First, we head off to investigating the single particle components of
the wave function. In the model versions which take into account the
backward amplitudes we found a serious depletion of the quasiparticle
strengths as exemplified in Fig.\ref{fig:fragmentation} for the case
of the qp state $\nu$2d$_{5/2}$. We found a similar behavior for the
rest of the states from the valence shell in all nuclei within the
considered region. An appropriate experimentally measurable quantity
to study the single particle strength is the spectroscopic factor (SF)
for the (d,p) reaction calculated as:

\begin{equation}
  S_{J\nu} = (C_{J\nu}u_{J} -  E_{J\nu}v_{J})^{2}
\end{equation}

From Fig.\ref{fig:sf1} we see that the value of $\omega_{2_{1}^{+}}$,
at which the experimentally measured spectroscopic factor is
reproduced, is lower in the case of EQPM$\_$BCK$\_$P than in
QPM$\_$BCK$\_$P by about 50keV and is therefore closer to the energy
of the first $2^+$ state in $^{130}$Ba. In Table \ref{tbl_SF}, this
comparison between the two model versions is extended for several
nuclei where experimental data are available. From there we see a
systematic improvement with respect to $\omega_{2_{1}^{+}}$ ranging
from 50keV to 150keV in favor of EQPM$\_$BCW$\_$P.

\begin{table}[b]
  \begin{center}
    \caption{ Spectroscopic factors for the $(d,p)$ reaction of the
      state $3/2^+_{1}$ in $^{123}$Te, $^{125}$Te, $^{127}$Te and
      $^{131}$Ba . The second column gives the experimental
      \cite{ENSDF} values. The columns 3 and 4 give the energies
      $\omega_{2_1^+}$ (in MeV) of the corresponding even-even cores
      calculated within QPM and EQPM at which the experimental values
      of the SF are reproduced.}
    \begin{tabular} {|c|c||c|c|c| } \hline
       Nuclide & Exp & {$\omega_{2_1^+}$, \small{QPM$\_$BCW$\_$P}} & {$\omega_{2_1^+}$, \small{EQPM$\_$BCW$\_$P} }  \\  \hline \hline
        $^{123}$Te & 0.5   & 1.4  & 1.3  \\
        $^{125}$Te & 0.46  & 1.5  & 1.3   \\
        $^{127}$Te & 0.38  & 1.5  & 1.35 \\
        $^{131}$Ba & 0.25  & 1.05  & 1  \\
      \hline
    \end{tabular}
    \label{tbl_SF}
  \end{center}
\end{table}


While the spectroscopic factors are influenced mainly by the
properties of the last, unpaired particle, the electric transition
probabilities depend strongly on the bulk properties of the even-even
core. The largest contribution to these quantities is due to
transitions between pure qp and qp$\times$ph states represented by the
sum in the rhs of the following expression:

\begin{figure*}
  \includegraphics[width=\textwidth]{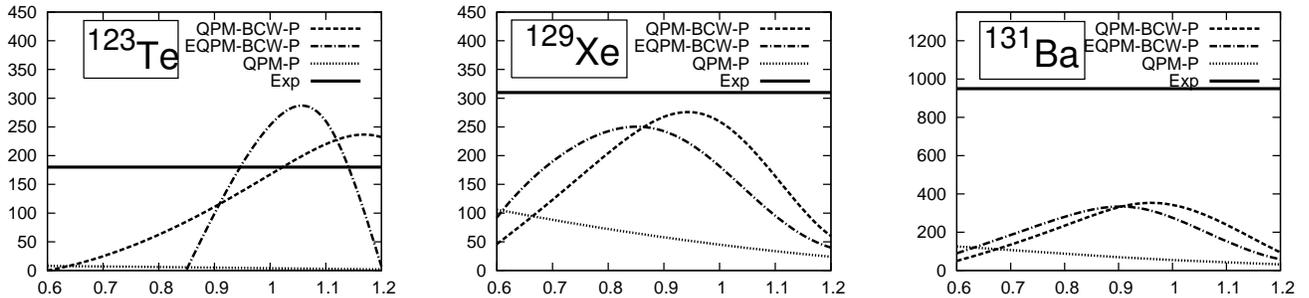}
  \caption{ \label{fig:odd_transitions} Same as Fig. \ref{fig:sf1} but
    for $B(E2|3/2_{1}^{+} \rightarrow 1/2_{1}^{+})$ in $^{123}$Te,
    $^{129}$Xe and $^{131}$Ba.}
\end{figure*}

\begin{widetext}
 \begin{equation}
   \label{eq:b_odd}
   B_{odd}(E\lambda ; J_1 \nu_1 \rightarrow J_2\nu_2) = \frac{1}{\pi_{J_1}^2}  \left(  C_{J_1\nu_1} C_{J_2\nu_2} e_{np} f_{J_1J_2}^{\lambda} v_{J_1J_2}^- + \sum_{i} U(J_{1}\nu_{1}J_{2}\nu_{2} \lambda i)  \sqrt{B(E\lambda ; g.s. \rightarrow \lambda_i)}  \right)^2,
 \end{equation}
 
 where $e_{np}$ is 1 if the unpaired particle is a proton and 0 - if
 neutron; $B(E\lambda ; g.s. \rightarrow \lambda_i)$ is the reduced
 transition probability in the corresponding even-even nucleus given
 by formula \eqref{eq:transition_even_even} and
 
 \begin{align}
   &U(J_{1}\nu_{1}J_{2}\nu_{2} \lambda i) =  \frac{\pi_{J_1}}{\pi_\lambda} [C_{J_2\nu_2}D_{J_2\lambda i}(J_1\nu_1) - E_{J_2\nu_2}F_{J_2\lambda i}(J_1\nu_1)][1 + L(J_1 J_2 \lambda i)]  + \nonumber \\
   &(-)^{J_1 - J_2 + \lambda} \frac{\pi_{J_2}}{\pi_{\lambda}}  [C_{J_1\nu_1}  D_{J_1\lambda i}(J_2\nu_2) - E_{J_1\nu_1}  F_{J_1\lambda i}(J_2\nu_2)] [1 + L(J_2 J_1 \lambda i)].
 \end{align}
 
 In expression \eqref{eq:b_odd} the terms corresponding to transitions
 between pure qp$\times$ph states have been neglected as being
 small. In systems where the last particle is a neutron we make the
 approximation:
 
 \begin{align}
   &B_{odd}(E\lambda ; J_1 \nu_1 \rightarrow J_2\nu_2) = \frac{1}{\pi_{J_1}^2}  \left[ \sum_{i}  U(J_{1}\nu_{1}J_{2}\nu_{2} \lambda i) \sqrt{B(E\lambda ; g.s. \rightarrow \lambda_i)}\right]^2 \approx \frac{1}{\pi_{J_1}^2}  U^{2}(J_{1}\nu_{1}J_{2}\nu_{2} \lambda 1) B(E\lambda ; g.s. \rightarrow \lambda_1) ,
 \end{align}
\end{widetext}

which stems from the fact that the coefficients
$U(J_{1}\nu_{1}J_{2}\nu_{2} \lambda i)$ are non-negligible for the
lowest-lying states only and out of these states the transition to the
first excited state is the strongest.

The dependence $B_{odd}(E2|3/2_{1}^{+} \rightarrow 1/2_{1}^{+}) =
B_{odd}(E2|3/2_{1}^{+} \rightarrow 1/2_{1}^{+})(\omega_{2_{1}^+})$ is
plotted in Fig. \ref{fig:odd_transitions} within the three model
versions. This function shows an almost linear behavior in the case of
the QPM$\_$P while in the calculations which take into account the
backward amplitudes a peak emerges. This peak is a result of the
increased fragmentation in the latter pair of model versions
(cf. Fig.\ref{fig:fragmentation}) which contributes to the enhanced
values of the coefficients $U(J_{1}\nu_{1}J_{2}\nu_{2} \lambda i)$. As
a result, the maximum value of the presented transition probabilities
in the EQPM$\_$BCK$\_$P and QPM$\_$BCK$\_$P is about three times as
large as the maximum value obtained within the QPM$\_$P bringing us
closer to the experimental values.  It is also worth noting that the
values of $\omega_{2_{1}^{+}}$, which correspond to the peak values
obtained within the EQPM$\_$BCK$\_$P, are about 100keV lower than in
the QPM$\_$BCK$\_$P. We therefore conclude that the effect of the
renormalization yields better results with respect to the
experimentally measured energy of the corresponding even-even core
though it is still rather higher from it.


\section{Summary and outlook}
\label{sec:conclusions}
In this work we extended the model presented in \cite{Mish08} by
taking into account the blocking effect due to the Pauli principle,
following the approach prescribed in \cite{Hara} and
\cite{Karadjov}. Renormalized quasiparticle-phonon interaction
strengths in both the forward and backward directions have been
derived. Numerical calculations on the spectroscopic factors and
transition probabilities in several Te, Xe and Ba isotopes have been
performed using a Wood-Saxon potential well and residual interaction
of a pairing+quadrupole type. The results indicate an overall improved
description of these experimentally measured quantities due to the
weakened quasiparticle-phonon interaction strengths.

However, despite the adoption of this elaborate approximation, further
improvements of the theory towards weakening of the qp-ph interaction
could resolve some of the existing discrepancies with the
experiment. Some steps in this direction would be the inclusion of
higher multipolarities, the use of multi-phonon configurations and the
development of a more elaborate approach to account for further
correlation effects.We finally conclude that our understanding of the
properties of the lowest-lying states in relatively stable odd-A
nuclei still lacks the desired accuracy \cite{2010_Bortignon}.


\end{document}